# Existential Ontology and Thinging Modeling in Software Engineering

Sabah Al-Fedaghi
Computer Engineering Department
Kuwait University
Kuwait
sabah.alfedaghi@ku.edu.kw

***Abstract*—This study is a sequel to a previous study entitled "Thinging for Software Engineers", which showed that the notion of *thing*, in contrast to objectification, has some beneficial orientations in modeling. The incorporation of thinging in conceptual modeling is required to explain the roots of Heidegger's conception of things. This requires an understanding of Heidegger's existential ontology to identify any relationship to thinging. This paper is an exploration of existential ontology in search of further clarification of the concept of thinging. We start by reviewing the thinging machine (TM) introduced in "Thinging for Software Engineers" and provide a full example of its utilization in modeling an "ordering system". We follow this with a discussion of the being (existence) of things in the word and Heidegger's interpretation of time as a possible horizon for any understanding whatsoever of being. We emphasize that the TM is not related directly to the Heideggerian notion of existence and its elaborate analysis of Dasein. However, there may be some benefit to studying non-Dasein things to provide a philosophical foundation to thinging, as utilized in TM modeling. Interestingly, the TM can be utilized to model existential ontology, thus increasing the level of understanding about them.**

***Keywords-conceptual modeling; thing vs. object; thinging; diagrammatic representation***

## I. Introduction

This study is a sequel to a previous study under the title "Thinging for Software Engineers", which was published in this journal [1]. Our motivation for that paper was the repeated rejection of submitted papers in conceptual modeling in software engineering based on, in the referees' words, "the use of vague terms such as things." "Thinging for Software Engineers" [1] shows that the notion of thing has a philosophical foundation in Heidegger's works, specifically in in his work entitled "The Thing" [2].

Our incentive for this paper was raised by students who are modeling using the thinging machine (TM) and the impression that that the notions thing and thinging have an origin in Heidegger's other works; thus, according to students, "to incorporate thinging in conceptual modeling in software engineering, we need to understand Heidegger's existential ontology which is very difficult if not impossible to comprehend by us." This paper is an attempt to find the roots of thinging in Heidegger's existential ontology. This would clarify further direction by providing a philosophical foundation to the TM approach.

It should be clear that this work is an exploratory explanation of existential ontology and that there is no claim of authority in the subject matter; however, the author must risk incurring disapproval, as there is no other available understood explanation of existential ontology for software engineers and, possibly, for most non-philosophers. Hence, if the interpretation of Heidegger's ideas is wrong, the presented materials can be considered the author's thoughts inspired by his reading of Heidegger.

This paper falls within the intersection of two research disciplines:

- Software engineering/modeling/conceptual
- Philosophy/ontology/existential

Although interdisciplinarity is fashionable in academia, it is typically viewed as cooperation among experts in the different disciplines. The work in this paper is characteristically not appreciated by specialists in both disciplines, as philosophy specialists consider it an intrusion on their turf and software engineering specialists think that it is not a practical effort and will lead to nothing. So, this paper needs (e.g., reading, refereeing) experts and generalists who are not strictly specialists. A specialist is a person who possesses special knowledge relating to a particular area of study. Generalists have an understanding of several subjects—in our case, conceptual modeling in software engineering and ontology in philosophy.

Conceptual modeling in software engineering is employed to facilitate, systemize, and aid the process of information engineering. Conceptual models describe entities of some domains in semantic terms [3]. The model serves as a tool for communicating between developers and users, thus helping analysts to understand a domain, providing input to the design process, and documenting purposes [3].

This paper adopts a conceptual model, called a thinging (abstract) machine (TM) that views all components of the domain in terms of a single notion: flow machine [4–12]. As our objective is to discuss the TM in terms of existential ontology, the TM is reviewed, and a full sample of the TM model is applied to a real system of ordering items.

Given the size of the research produced over the years on Heidegger's thoughts, another review of the literature is not necessary. Instead, simplified Heideggerian statements (e.g., no non-English terms) will be interwoven into the appropriate text in the paper. We assume that the reader is familiar with basic philosophical terminology (e.g., ontology) and software engineering (e.g., object orientation).

## II. THINGING MACHINE (TM)

We adopt the TM, a conceptual model that is built on handling (e.g., creating, processing) things [2] and machines (assemblage – unified gathering). The TM can be introduced without mentioning its philosophical base; however, in this case, the model may be criticized based on the alleged ambiguity of the term "things". Additionally, the TM can contribute to object-oriented modeling in dimensions, such as uniformity and continuity of description, as will be indicated later in this paper. Or, the TM can be used as one additional type of UML diagrams.

According to TM methodology, instead of perceiving things as objects (e.g., object-orientation), we conceptualize things in their assemblage, or the residence of things "invested with value" (of interest for modeling), where things *things* (verb). A book as a thing has "vast historical conditions" and "social contingencies" with all of its singular features (e.g., color, lighting, time of day) and all the conditions under which someone is inquiring into the book [13]. Note that a thing can be conceptualized as a machine and that a machine can be conceptualized as a thing.

From the Heideggerian perspective, things, each in their own way, are encountered as equipment, ready-to-hand or present-to-hand (Heideggerian's terms), as they are not just put to any specific use but are participants in the complex mesh of interwoven and interacting entities, creating a whole master machine. Equipment are things that we encounter in concern (Heidegger). We deal with things in a kind of concern that manipulates things and puts them to use.

In the TM, we capture this dealing with things as equipment through the notion of machinery. Things are manipulated by creating, processing, and transposing them among different machines. In contrast to Heidegger, whose concern is solely with the theme of being, our concern is with how to capture the manipulativeness and usage of things. We define a thing as what is created, processed, released, transferred, and received. Creation may be taken as the first stage of Heidegger's being-ness.

The simplest type of this thing/machine, according to our modeling approach, is called the TM, as shown in Fig. 1. Machines are the conceptual space of things and flows. They may reside in other machines, and this relationship of machines can be expanded until we can say that that all are inside a grand machine called the TM model of an organization.

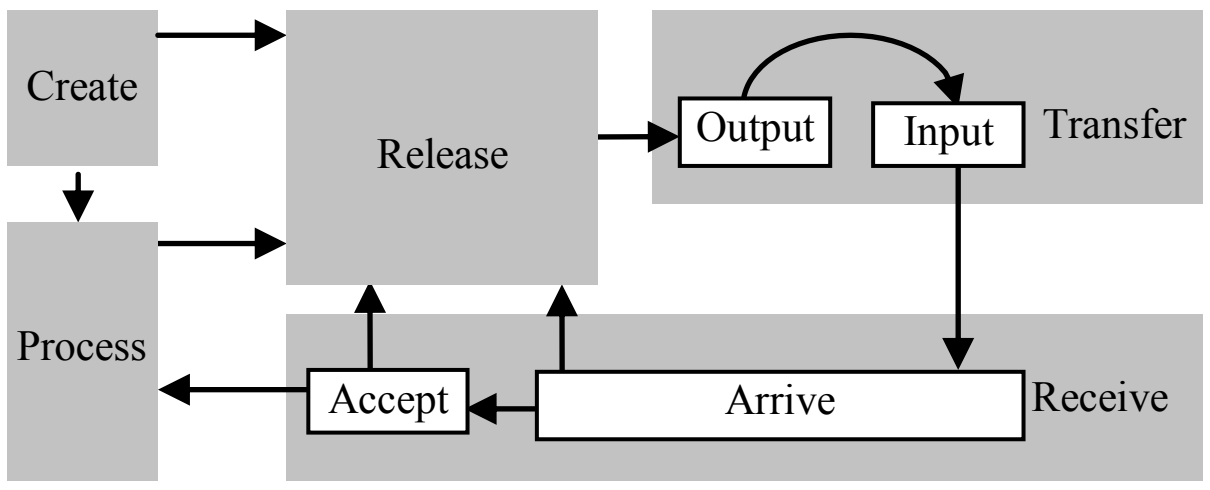

Figure 1. Thinging machine.

The arrows represent conceptual flows. Conceptual flow is not necessarily a physical flow. For example, in a manufacturing assembly line, if a device (thing) arrives at a position where two robots process two parts of the device, there are simultaneous flows (transfer, receive, and process) to the two robot machines (conceptual space).

Regarding the directions of the arrows in the TM, note that, philosophically, we follow the orientation of Heidegger's ready-to-hand [14]. This assumption in the TM leads to only forward flow. For example, suppose that an organization sends a package through the DHL Company, where the package is released and transfer to transfer and receive in the DHL. The assumption here is that DHL never reverses the flow (say, because of emergency stoppage of transportation), such that the package flows backward. This is a typical engineering assumption, as, in the case, say, of a model of crude oil distillation where a thing (e.g., crude oil) flows forward in a vertical fashion from one stage to another. In the model of such a process, the possibility that the oil may flow backward, in the case of a distillation unit breakdown (present-to-hand), is not included in the description. The TM handles the situation of DHL returning the package by creating a new thing (say, Returned Package) and releasing and transferring it back through a different flow than the original one. Accordingly, the TM arrows' direction remains intact and is used in both flows.

The TM creates, processes, receives, releases, and transfers things. The TM model is an overarching machine that forms the thinging of a system. Thinging involves "defining a boundary around some portion of reality, separating it from everything else, and then labeling that portion of reality with a name" [15]. To understand thingness philosophically, one needs to reflect on how thinging expresses the way a thing "things"—that is, gathers, unites, or ties together its constituents—in the same way that a bridge unifies aspects of its environment (e.g., a stream, its banks, and the surrounding landscape) [14].

In a TM, the flow of things refers to the exclusive movement of things among the five operations (stages) shown in Fig. 1. Although things can be stored in addition to being created, processed, released, transferred, and received, being stored is not a generic operation. For example, things can be stored after being created (thus becoming stored created data) or after being processed (thus becoming stored processed data), and so on. After all arriving things have been accepted, the combination of arrive and accept is represented by the term "receive".

A thing in a machine, in addition to being imported (transferred or received), can be created (what gets produced – Heidegger). As a machine, the TM becomes aware of new things via either creating or importing. Additionally, a thing disappears from a machine's view when it is either deleted or exported (released or transferred). Note that a thing can be released but not transferred (e.g., when finished goods are waiting for a truck to arrive before being shipped) or transferred without arriving (e.g., when an e-mail is sent but an error prevents the recipient from accessing it). A process (what gets used – Heidegger) occurs when a machine changes a thing in a certain way. For example, a doctor machine could process a patient to decide on the appropriate treatment. Note that release, transfer, and receive are not Heidegger's notions.

Each type of flow is distinguished from other flows. No two flow streams are mixed, just as telephone and water lines are separate in buildings' blueprints. However, two types of things can enter a machine for a shared supertype (e.g., integers and real numbers flowing to a number machine). A TM does not necessarily include all stages (e.g., an archiving system might only use the transfer, receive, release, and process stages, leaving out the create stage).

Machines can interact through flows or by triggering new stages. Triggering is a transformation (denoted by **a dashed arrow**) from one flow to another (e.g., when a flow of electricity triggers a flow of air).

## III. Example of Thinging Machine Modeling

This paper is about thinging and explores strengthening its foundation. Hence, to have a self-contained paper, we provide a full example of the TM in this section.

According to Visual Paradigm [16], a state diagram is a type of UML behavior diagram. It describes all of the possible states of an object (or even an entire system) and provides the means to control decisions. The object behaves differently depending on its state. A simple ordering system is described in terms of state, and activity diagrams are given, as shown in Figs. 2 and 3, for an ordering system.

In UML semantics, activity diagrams are reducible to state machines, with some additional notations to indicate that the vertices represent an activity being carried out; the edges represent the transition from the completion of one collection of activities to the commencement of a new collection. Activity diagrams capture the aspects of high-level activities [16].

As this ordering system is already modeled using UML, this provides an opportunity to contrast it with the alternative TM modeling.

### A. *Static Thinging Machine Model*

The initial TM diagrammatic description is static in Heidegger's sense of a present-at-hand multiplicity of "nows", as will be explained later. Fig. 4 shows the ordering system, as understood from the state and activity diagrams. In Fig. 4, an ordered list is created (1) in the lower left corner of the figure. The order list flows to the ordering system (2), where it is stored (3) and processed (4) to create an invoice (5).

The invoice then flows to the customer (6). Sending the invoice to the customer triggers the setting of the deadline (7) for receiving the payment; if the payment is not received on time (8), lateness triggers the deletion of the order (9). If the customer creates a payment (10) that the ordering system receives (11), the following activities are the result:

- A trigger stops the timer for payment (12).
- A trigger extracts (13) items (14) from the list. Note that the ordered items include two components: the items' names and the number of ordered items (15).

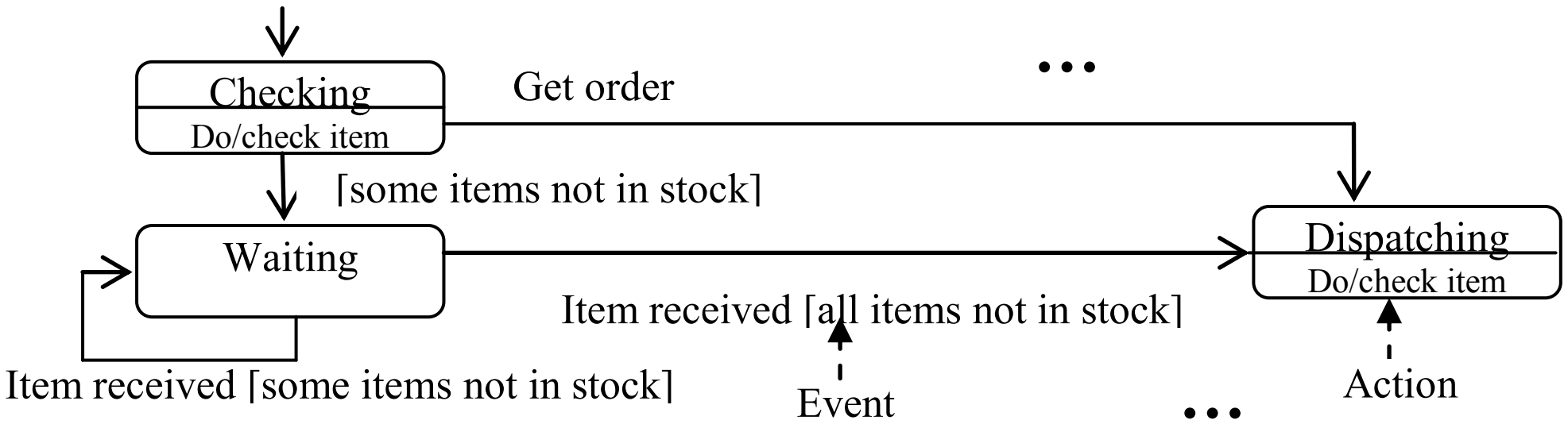

Figure 2. State diagram for ordering items (Redrawn, partial from [16]).

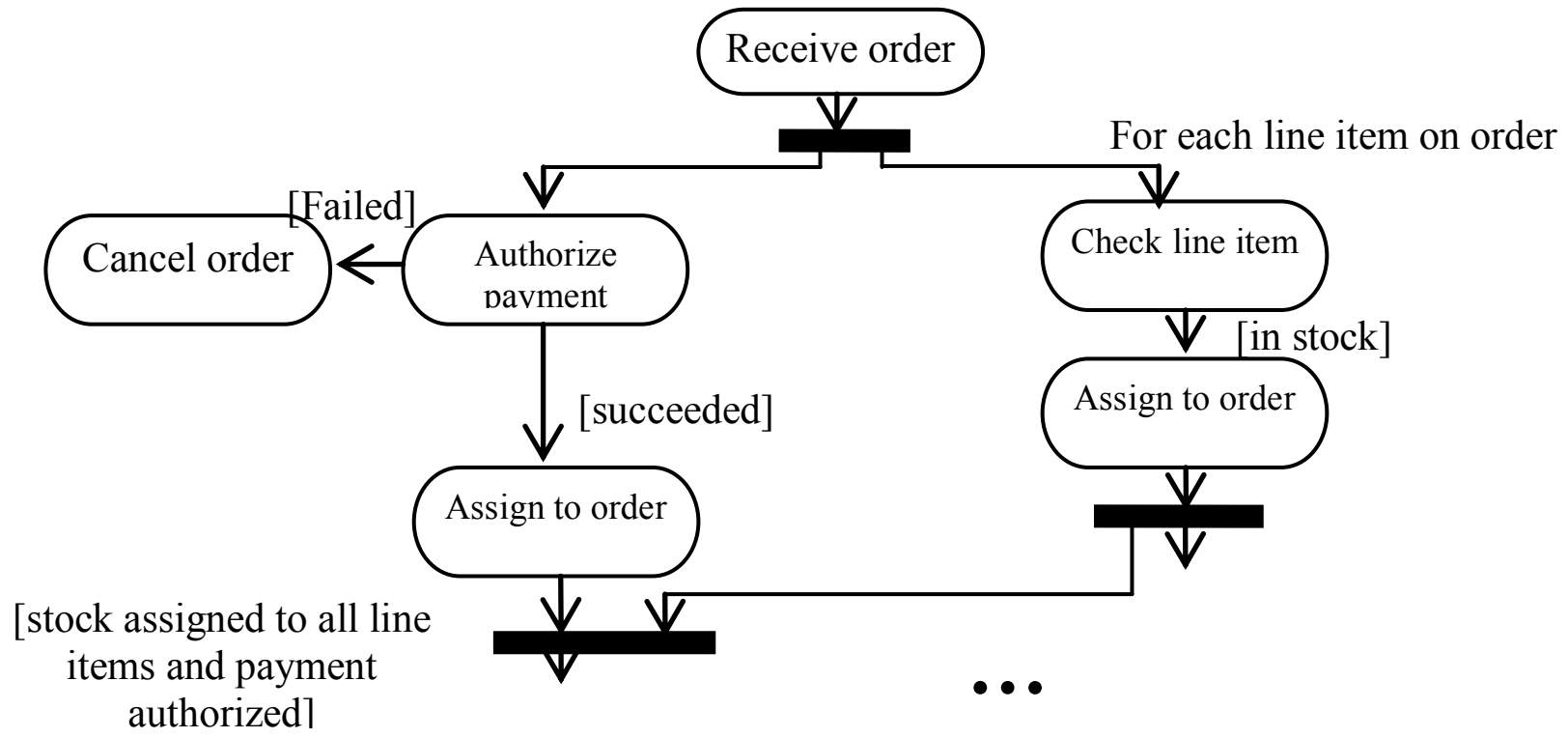

Figure 3. Activity diagram for ordering items (Redrawn, partial from [16]).

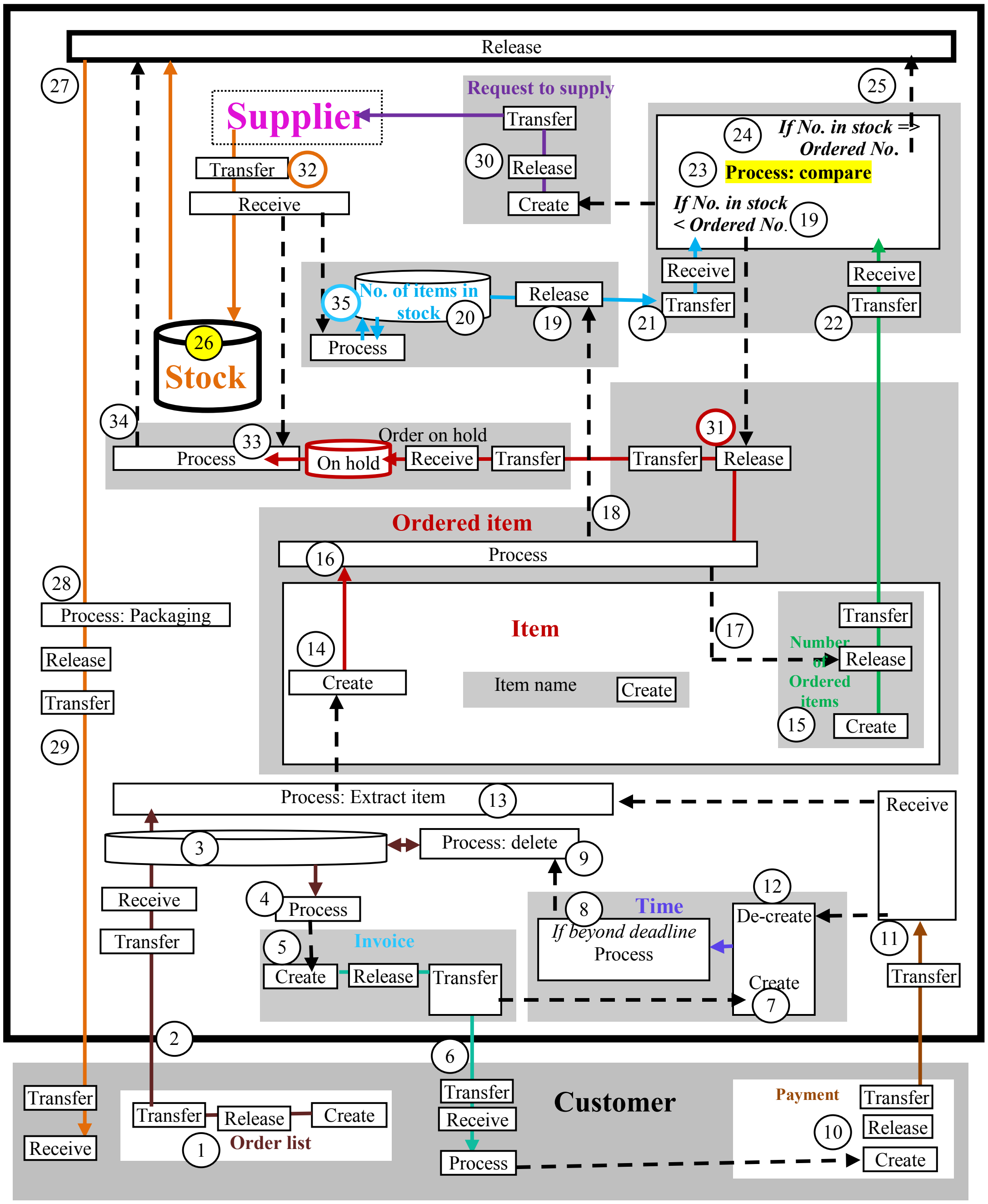

Figure 4. Thinging machine diagram for ordering items.

Processing the ordered items (16) results in the following:

- The number of ordered items is extracted (17).
- Simultaneously, a trigger (18) releases (19, middle of the figure) the current number of items in stock (20).

The above two values are sent (21 and 22) for comparison (23):

- If the number of items in stock equals or is greater than the number of ordered items (24, upper right corner), the ordered items are released (25) from stock (26, yellow circle) and flow (27) to the packaging process (28). They are then sent to the customer who made the order (29, lower right corner).

If the number of items in stock is less than the number of ordered items, two things can happen:

(a) A request is issued for the supplier to provide new items (30).

(b) The order is put on hold (31, red circle).

When a requested item is received (32, orange circle), it is put in stock (26, yellow circle) and then followed by the subsequent steps:

- The on-hold item is processed to trigger (33) the release (34), the packaging (28), and the sending (18) of the item to the customer who ordered it.
- The number of items in stock is updated (35, blue circle).

## B. *Events and time*

To specify the behaviors in such a system, we identify the sequence events in the diagram. Every change in the model can be considered an event. However, at the practical level, these events are combined to form different combinations of events. To illustrate, we outline one of these combinations as follows (see Fig. 5):

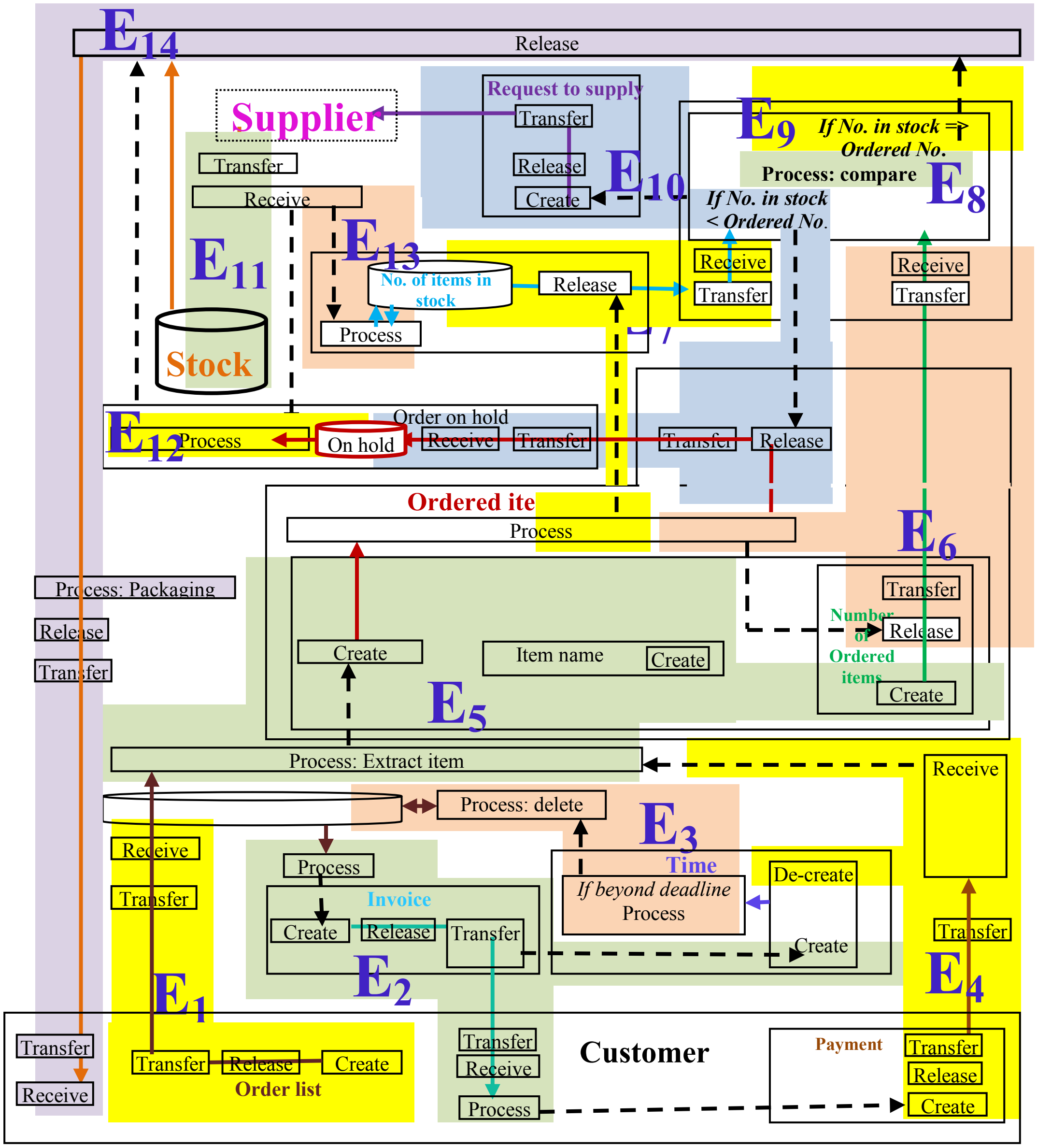

Figure 5. Events.

Event 1 ($E_1$): An order is received.
Event 2 ($E_2$): An invoice is sent, and the payment deadline is set.
Event 3 ($E_3$): After the deadline has passed, the order is deleted.
Event 4 ($E_4$): Payment is received.
Event 5 ($E_5$): The item is extracted from the list.
Event 6 ($E_6$): The item is processed.
Event 7 ($E_7$): The number of items in stock flows to the next step to be compared.
Event 8 ($E_8$): The number of items in stock is compared.
Event 9 ($E_9$): The required ordered number is available in stock.
Event 10 ($E_{10}$): The required ordered number is not in stock, so a request for more supplies is sent to the supplier, and the order is put on hold.
Event 11 ($E_{11}$): The requested supplies arrive.
Accordingly, the chronology of events is shown in Fig. 6.

The TM model provides a rich diagrammatic environment that improves, accompanies, and can replace some current, compact diagrams. Its conceptual description makes it suitable for applications in many fields. It captures the description of a portion of the world that "exists" and provides a base for building a software system. The flows inside the model, physical or otherwise, form a fabric that interweaves its components to make the representation of that portion of the world emerge. Activities are viewed as the *machine* of *things* that flows through that fabric.

The resultant conceptual description is Dasein (human)-made theoretical constructs of a certain universe of activities. The construction explains a machine (the grand diagram), its slices and how its things behave, and their flows and submachines that are inside a definite enclosure or in the Heideggerian language a certain gathering. In ordinary ontology, unconceptuality indicates raw, unformed, and unshaped content (e.g., passive sense-data) or given bare presence (given physical materials). According to [17], "a model is an abstraction of something for the purpose of understanding it before building it."

The philosophical foundation of this modeling has been based on Heidegger's notion of thinging. To strengthen such a foundation, we have to explore Heidegger's existential ontology.

## IV. EXISTENTIAL ONTOLOGY

This section is about existential ontology, not for the purpose of contributing thing in this field; rather, our aim is to explore how it supplements thinging. In philosophy, Heidegger's existential ontology is concerned with the being (existence) of an entity in the word and "the Interpretation of time as the possible horizon for any understanding whatsoever of Being" [14]. Ontology refers to the meaning of entities. For Heidegger, the "task of ontology is to explain Being itself and to make the Being of entities stand out in full relief" [14]. According to Dreyfus and Rabilnow, "Heidegger's existential ontology is the best description of human social being that philosophers have yet offered, but it is totally abstract" [18].

The Heideggerian explanation of being is framed in the way we have come to terms with *the things themselves* (phenomenology – the science of phenomena). According to Heidegger (as we understand it), being is not a class or genus of entities and cannot be conceived as an entity. Being cannot be derived from higher concepts, nor can it be presented through lower ones.

As an initial attempt in our understanding of this being, we conceptualize being in term of a *messy blanket* that extends infinitely as an ocean of waves (entities), as shown in Fig. 7. In this case, being is messy blanketing.

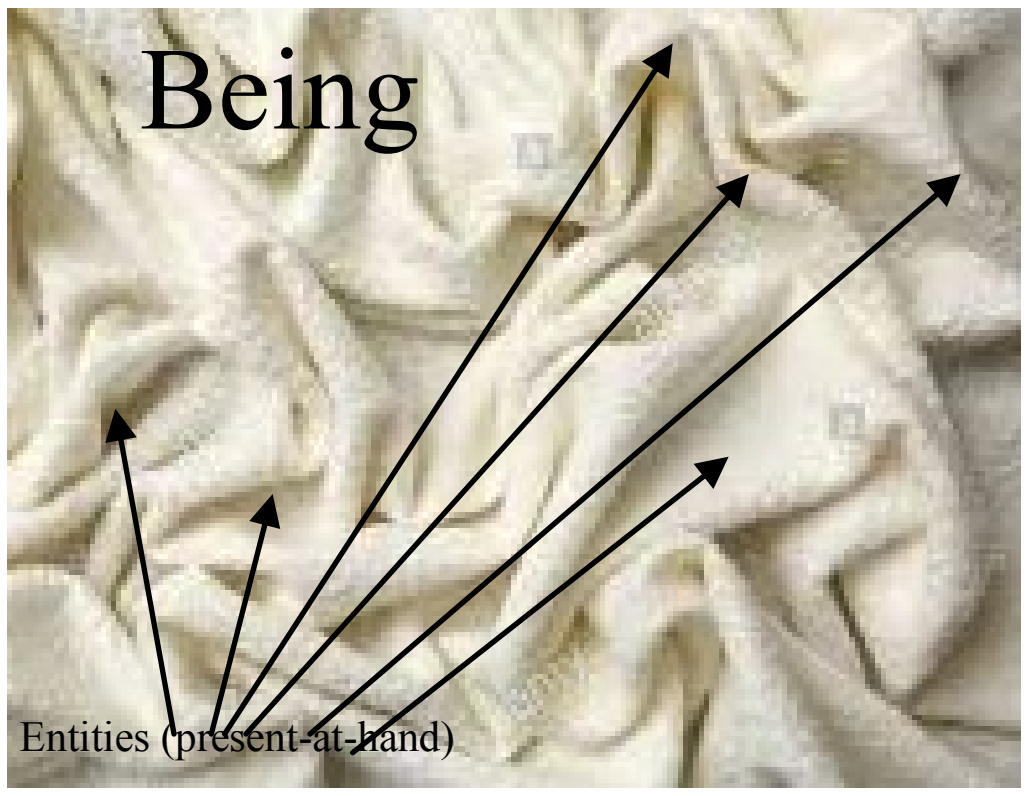

Figure 7. Being and its entities in terms of a messy blanket.

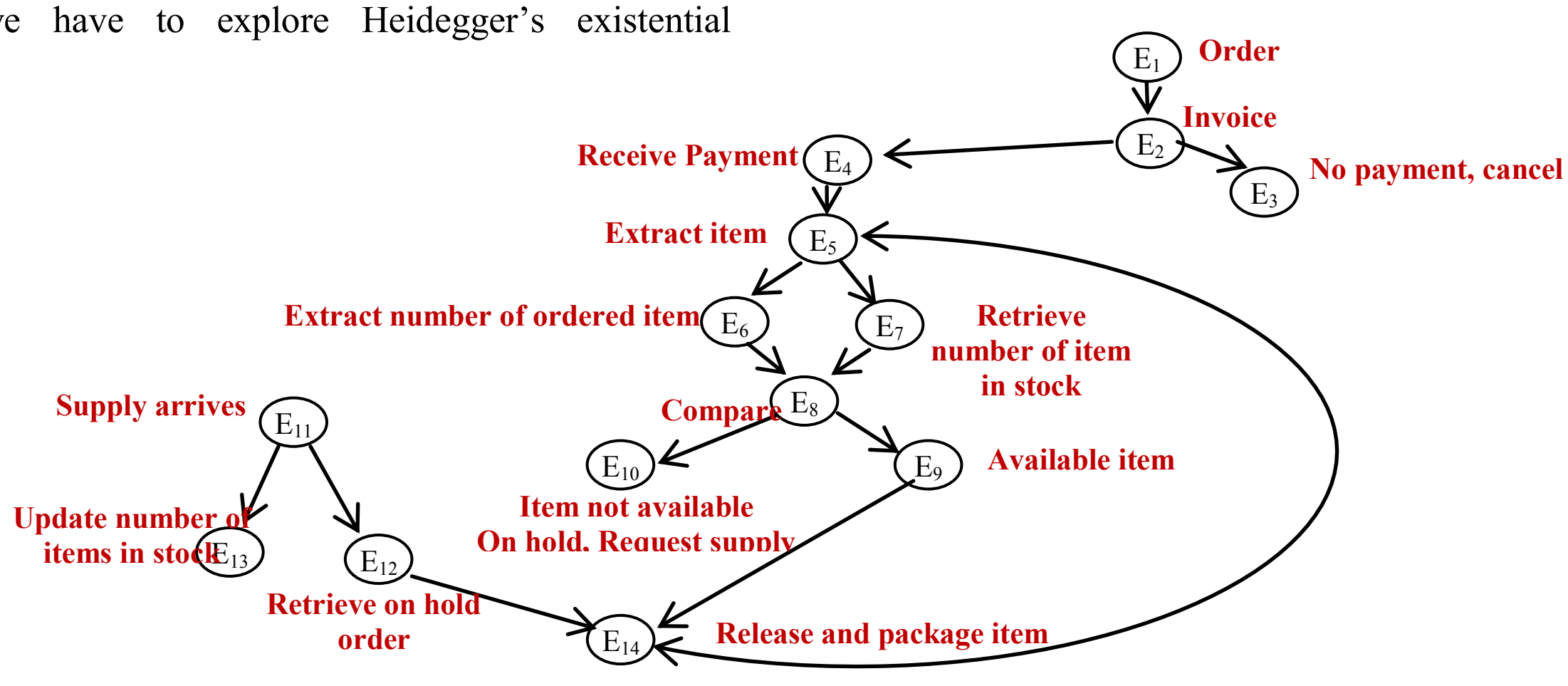

Figure 6. The execution of events.

Additionally, much like ocean waves, the messiness of this blanket changes over time, as some waves remain as they are or change their shape and others disappear and are replaced by new entities. Some [live] entities produce [little] waves that grow to become full waves, and some move as nomads over the blanket. From a TM point of view, we are most interested in these waves, but Heidegger's interest is directed at something else.

A special type of being is called Dasein. According to Heidegger, the "essence" of this entity lies in its "to be". Dasein is a different type of being-in-the-world than blanketing and its waves, which is characterized by a type of blanketing called existence. The traditional term "existential" is equivalent to being-present-at-hand, which is a kind of being inappropriate to entities of Dasein's character (Heidegger).

Dasein's characteristics are ways for it to be. However, Dasein is not some kind of ontologically independent that stands apart from, and above, the stream of changing experiences of the wave [19]. Dasein knows its blanket (e.g., smooth and rough parts), make uses of that knowledge, and recognizes and uses encountered blanket waves. In a TM, as we mentioned previously, a thing (wave) is a machine and a machine is a thing. The Dasein is a machine that "processes" in a special way (e.g., understanding) the blanket (see Fig. 8) and its waves.

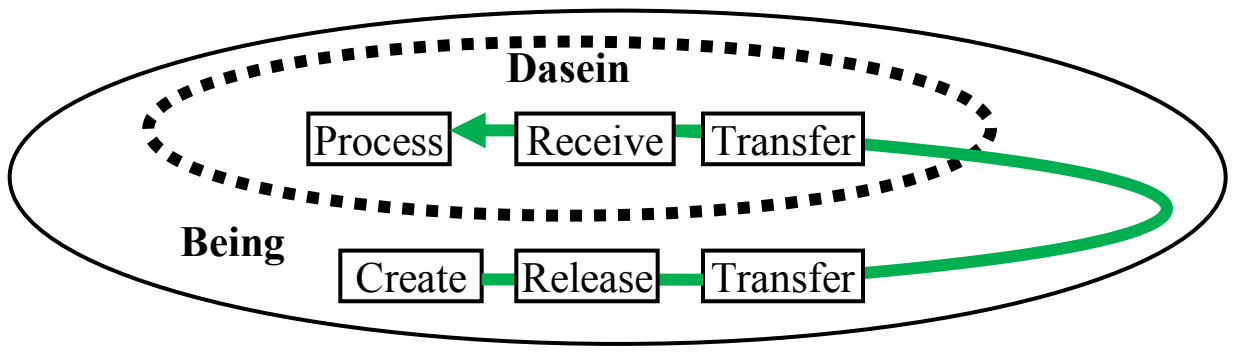

Figure 8. The Dasein processes being.

According to Heidegger, "Dasein, in its Being, has a relationship towards that Being [the blanket], a relationship which itself is one of Being." That seems to mean that the Dasein has a "wavy" relationship with the blanket. It is not a wave but instead has its own waves (body) and can create waves (e.g., concepts), process waves, etc. It is not a thing (wave), a substance (blanket materials), nor an object (what this blanket discussion is directed at). It is more than a mere piece of blanket (present-at-hand) endowed with special characteristics (e.g., intelligence, life). Note that the TM seems to be suitable for expressing these ideas diagrammatically.

According to Heidegger (and our understanding), Dasein is created (born) "in" the blanket (thrown in) at a certain time to dwell (falling in) in it, and at a certain point in time, his/her relationship with the blanket is "terminated". He/she does not "know" why he/she was created in the blanker, the purpose of the dwelling, and the moment of being terminated. He/does not know if he/she was anything before being in the blanket or if he/she will be anything after ending the Dasein being in the blanket.

From a modeling (diagrammatic) viewpoint, the big question, at this point, is *where* is the Dasein in the blanket? Is it a special type of wave? This is seemingly not possible because, as it is a wave in the blanket, the blanket has a "soul". Let us examine again what Heidegger writes about this Dasein:

- Dasein has priority over all other entities in terms of understanding and discovering the being of all other entities, both in the fact that they are, as well as in their being.
- Dasein characteristics are ways for it to be.
- All the being-as-it-is that the Dasein possesses is primarily being.
- Dasein is an entity whose being has the determinate character of existence.
- Dasein understands its own being (develops or decays with time) in terms of that entity toward which it comports itself proximally and constant-in terms of the world.
- The very asking of this question [of being] is Dasein's mode of being, and, as such, it gets its essential character from what is inquired about, namely, being.
- This entity in which each of us is himself.
- Dasein is an entity in which I myself am for each case.

**Heidegger: "Dasein is we are it."**

The blanket has reached its limits as a pretense of the totality of being. The Dasein cannot be incorporated in the blanket ontology. We have to *project* (i.e., to cast, conceive, extrapolate) the blanket on being-in-the-world, the "real" being, maintaining the notion of being as a unitary phenomenon.

This results in a real world that replaces the blanket and incorporates the Dasein in this world: We are **it**—I (the author) and you (the reader) who accompanied me in this inquiry about being. In this case, the blanket is replaced, project-ively (same structure) to arrive at us, the Dasein in the totality of being. Thus, the blanket disappears into our being, and the waves are the things of our being, as illustrated in Fig. 9. Note that we (the author and you the reader) are the creators of our blanket as being. I am the being whose essence lies in my "to be", as shown in Fig. 10. I am a constitutive part of the figure.

As I (the author) have discovered that Heidegger is embedding me in his description of being, I am the maker (Dasein's priority), a constitutive part of the blanket, and the way in which my modes of existence are in each case mine. I can go back to the blanket as being without the risk of imposing the Dasein as a piece of the blanket, which is an analogy that the blanket cannot incorporate but that real being can. I can consider myself a wave in the blanket without creating confusion.

The temporality of the Dasein is another important notion. To understand it, we look at some of Heidegger's points on this.

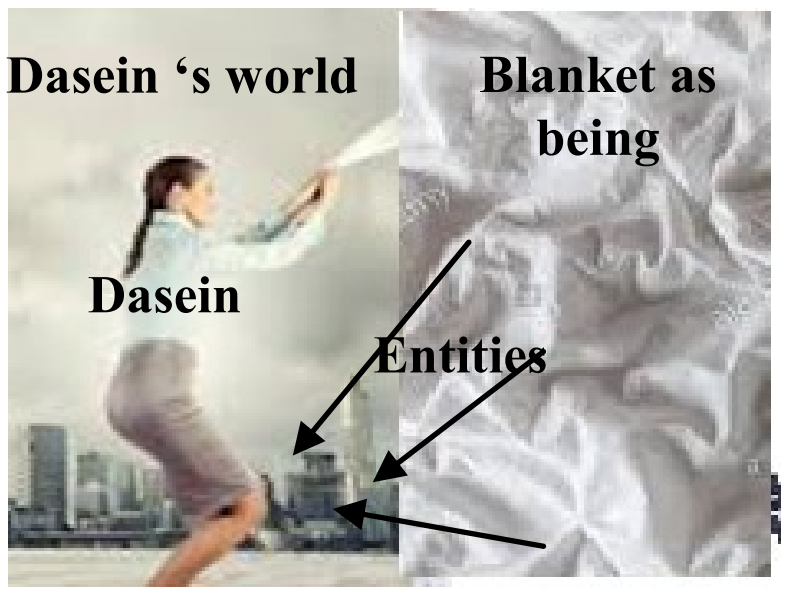

Figure 9. Projecting the blanket to reality to insert the Dasein into being

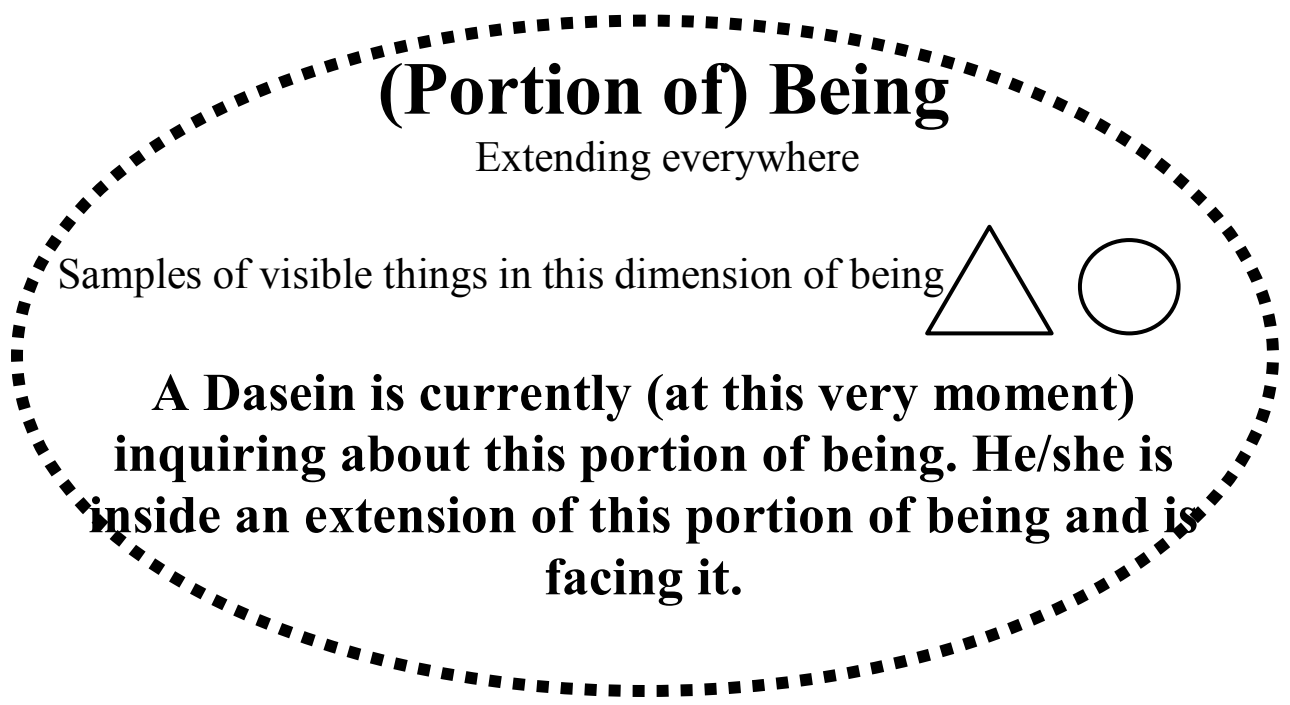

Figure 10. The Dasein is the being whose essence lies in its "to be".

- Temporality is the meaning of the being of Dasein. Temporal means being in time, and this functions as a criterion for distinguishing realms of being.
- Being is made visible in its temporal character.
- Structures of Dasein are interpreted as modes and derivatives of temporality.

Accordingly, as a wave in the blanket (Dasein's "own understanding of Being"), Dasein will end in death at a certain point in time. The disintegration of the Dasein means the disappearance of the whole of its blanket because the Dasein (I) is the creator of that being. So death, from my perspective, is the death of my world of being. The "lights turn off" and the Dasein and his/her world turns into nothing. Does this mean that the world is imaginary? No, and this is where the focus of the paper the word of things.

## V. EXISTENTIAL ONTOLOGY: THINGS

In this section, the subject matter is understanding the notion of being-in-the-world as a unitary phenomenon. As we saw in the blanket analogy, being-in-the-world cannot be broken up into content that can be pieced together (Dasein and non-Dasein). A thing as being-in-the-world manifests or unfolds itself in our being space. It is even possible for an entity to show itself as something that it in itself is not. [2]. The appearance of a thing does not mean showing itself. Things may show themselves and indicate something that does not show itself, as in the case of the symptoms of a disease. This is expressed in the TM, as shown in Fig. 11. Present-at-hand things outside of Dasein meet up with it, only in so far as it can, of its own accord, show itself within a world.

Thus, things show themselves to the Dasein who built his/her blanket. With the Dasein's death, the things of the Dasein blanket disappear, but the real things remain in their own being. At this point, we have clarified our objective to connect thinging with existential ontology.

Just as we conceptualize being in terms of a messy blanket, the TM diagram is a conceptualization of "concern" (i.e., to carry out something, to get it done, to straighten it out, to provide oneself with something) about a portion of being-in-the-world. Dasein being toward the world is essentially a concern. It is the Dasein's understanding of its own being. It is that result of perception (making determinate) to be expressed in diagrams.

However, the TM diagram is not about Dasein but about things in the world. It looks like a description of "mental representation" that is disliked by Heidegger, but we consider it, at this stage of development, an expansive abstract, which is a notion that encompasses the Heideggerian holistic view of *thing* and *thinging,* where thinging is an abstraction-like process that deemphasizes reduction and hence facilitates seeing the bigger picture instead of the reductive nature of *object*-oriented modeling.

The initial static TM diagram is, in Heidegger words, "a present-at-hand multiplicity of 'nows.'" Time orders these nows. For example, according to Heidegger, the sun, whose warmth is in its everyday use, "has its own places-sunrise, midday, sunset, midnight; these are discovered in circumspection and treated distinctively in terms of changes in the usability of what the sun bestows" [14]. Fig. 12 shows the TM's static (present-at-hand) modeling of this situation, where the sun creates warmth that flows to a region on Earth. Fig. 13 illustrates the notion of time in terms of an event where events change over the same region.

Consider the object-oriented approach to modeling. From the early stages of this paradigm's emergence, there were and still are many claims of the many benefits of the object-oriented approach, including the fact that the object-oriented system is distinguished by its potential capability of capturing the meaning of the application: its semantics [20]. Object is a fundamental notion in object-orientation.

An object-oriented model only includes things strictly serving the purpose at hand [21]. The modeler tries to identify the core concepts and sketches their relations and behaviors. Accordingly, what is a vehicle? It is a class of objects that are formed from data and methods, etc. This "conceptualization" needs no elaboration, as it is the bread and butter of software engineers. In such an ontology, "the thing itself is deeply veiled" (Heidegger) in computer technology, data structures, and programming.

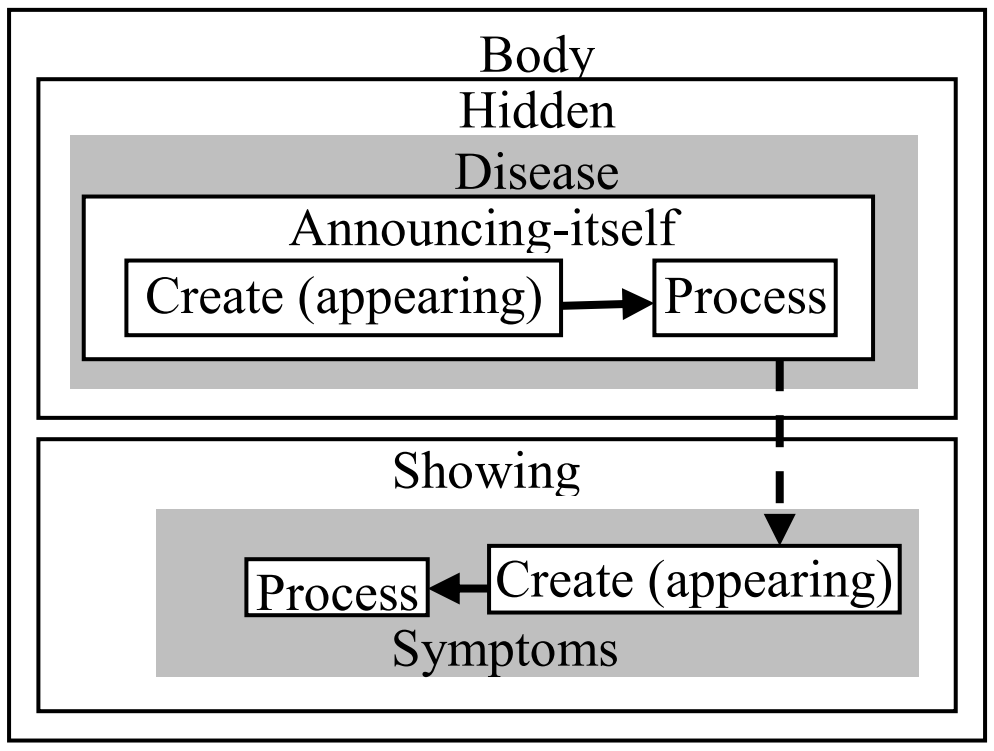

Figure 11. Things may show themselves and indicate something that does not show itself.

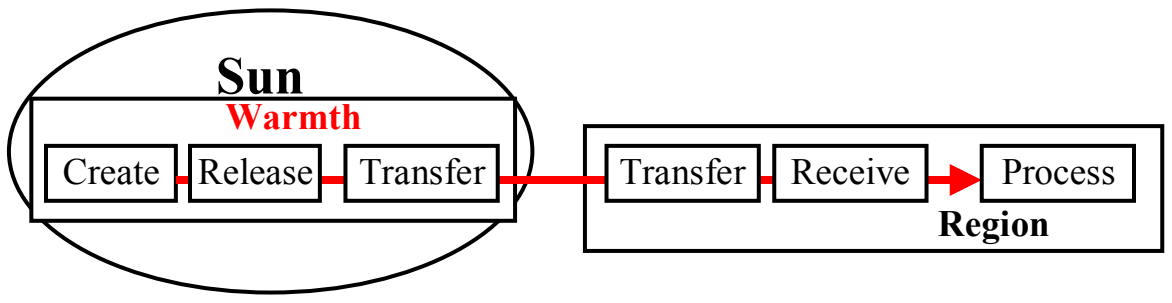

Figure 12. Illustration of a timeless thinging machine model (Adapted from [14]).

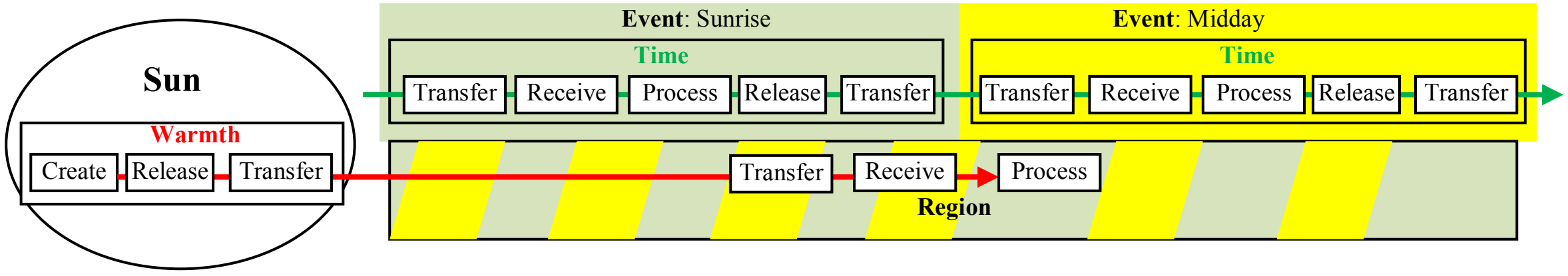

Figure 13. Illustration of the notions of events and region (Adapted from (Heidegger (being and time)).

The modeler–entity relationship in object orientation is illustrated in Fig. 14. An existential orientation toward entities is illustrated in Fig. 15. We propose that the TM, with its holistic-integrated methodology, can contribute to bringing these two orientations closer together. It is worth trying to strengthen its foundation in this paper.

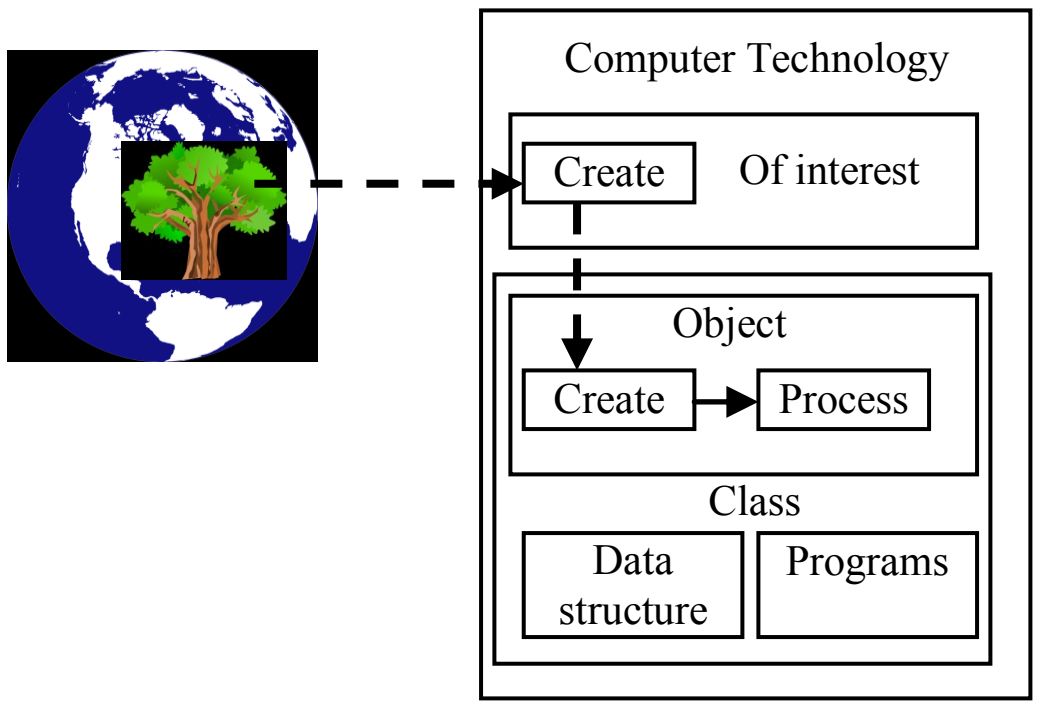

Figure 14. Object orientation.

## VI. Thinging Machine-based Exploration of Existential Ontology

This section is an attempt to schematize some of Heidegger's ontological concepts using the TM. After all, this paper is motivated by the complaints of computer engineering students who have tried to master the notion of thinging by studying existential ontology. According to Schwill, "it is necessary that students obtain a sketch of the fundamental ideas, principles, methods and ways of thinking . . . Only these fundamentals seem to remain valid in the long term and enable students to acquire new concepts successfully during their professional career" [22]. Because of space limitations, we only discuss a few notions of existential ontology to demonstrate the viability of the TM's diagramming method. The message here is to *think* Heidegger in terms of the TM. This would increase the level of understanding of both of them.

According to Heidegger, Dasein, as grounded in temporality, is in its very existing. The understanding and interpretation of both Dasein and time belong to existence. The human being has been thrown into the "there" of its being-in-the-world. It reveals itself as something that has been thrown. This line of thought may be beneficial for exploring the notion of time and thus clarifying the behavioral aspect of the TM.

Dasein has "time" itself in mind. The time is present-at-hand as an entity within-the-world (which it can never be) and because it belongs to the world in the sense of existential-ontologically.

Time is what is "counted". The nows are what get counted. And these show themselves in every "now" as nows that will "forthwith be no-longer-now" and nows that have "just been not-yet-now." Time is understood as a succession or as a "flowing stream" of nows. The nows are present-at-hand in the same way as things. The nows pass away, and those which have passed away make up the past. The nows come along, and those which are coming along define the "future". The time character has a location of the same kind as Dasein's.

This understanding of time is represented in Fig. 16. The lower stream of flow denotes the flow of Dasein.

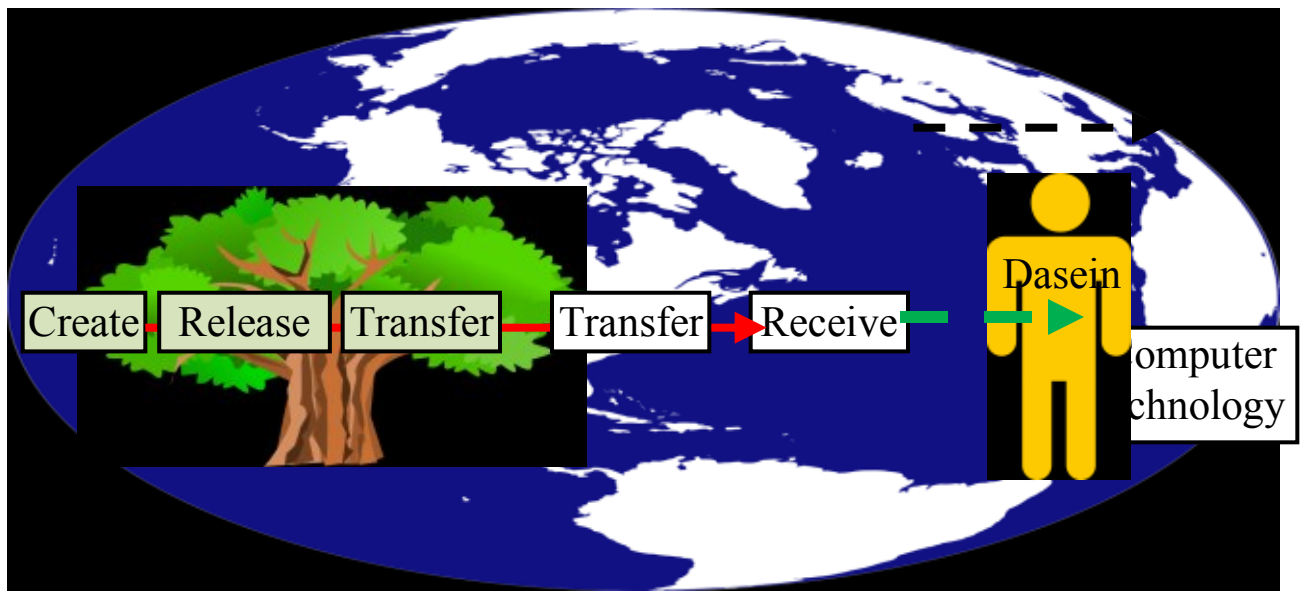

Figure 15. Existential orientation.

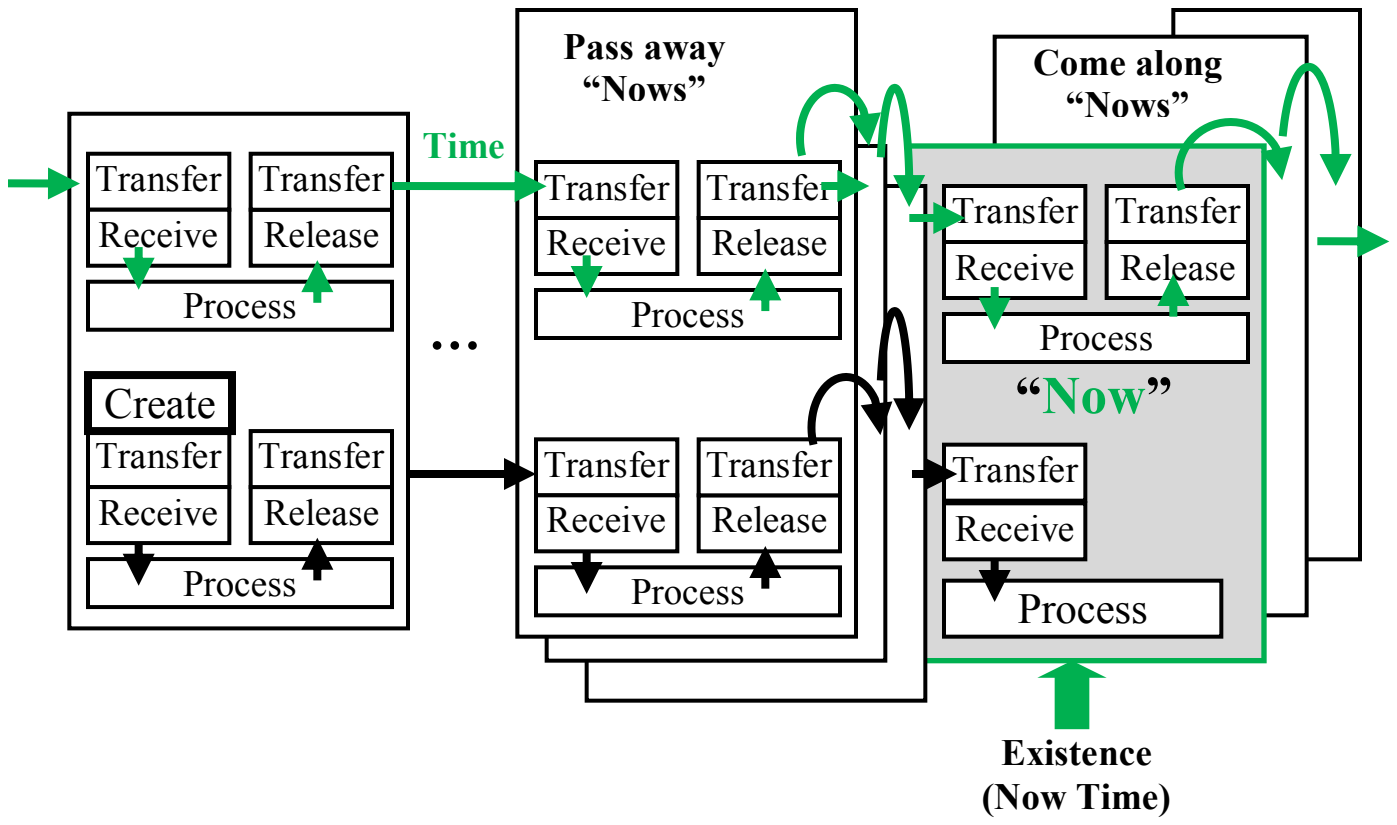

Figure 16. The nows come along, and those that are coming along define the "future".

The left most rectangle is the event of the Dasein appearance (create). Then he/she flows in being from one event to another. As mentioned above, the nows come along, and those which are coming along define the future. Because now is time, existence is defined in terms of time.

According to Wheeler [23] in the *Stanford Encyclopedia of Philosophy,* Dasein's three temporal dimensions are past (thrownness), future (projection), and present (fallenness). Dasein's existence is understood by way of an interconnected pair: thrownness-projection-fallenness and disposedness-understanding-fascination. These dimensions can be added, as shown in Fig. 17.

## VII. Conclusion

This paper started by reviewing thinging machines (TMs) so as to enhance the TM approach and strengthen its potential viability as a tool in conceptual modeling. It is a venture to explore the roots in Heidegger's conception of things. Accordingly, existential ontology is explained through examples and analogies.

The results do not seem to be satisfactory, as too much effort is spent in understanding the Dasein. Our goal is centered on thinging and not the Dasein. In existential ontology, Heidegger's whole focus is the Dasein, and his discussion of non-Dasein things is directed to serve this purpose. It is not clear how to concentrate mainly on non-Dasein things, as Dasein is the only one who understands being and its own being. Perhaps this is what Heidegger was trying to do in the "The thing" [2]. Further works should proceed further in potentially reaching non-Dasein things as the stars of the show.

On the positive side, we demonstrated that the TM is applicable in illustrating and explaining existential ontology, thus strengthening the relationship between conceptual modeling and philosophy.

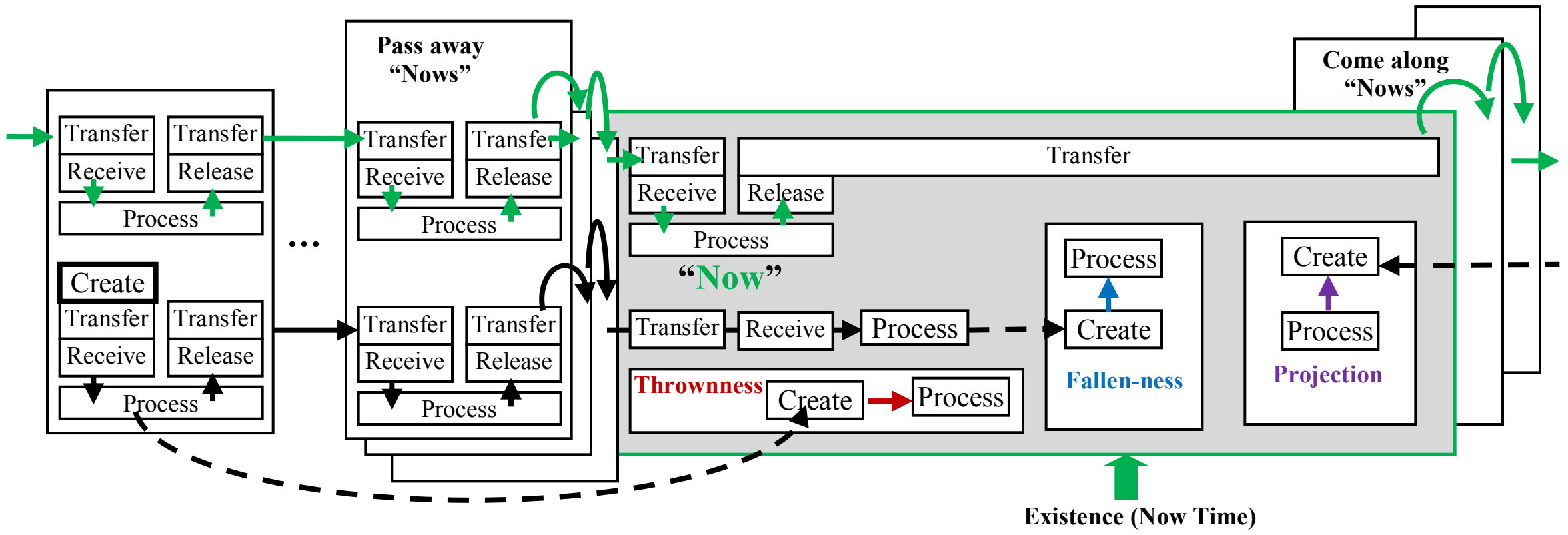

Fig. 17. Dasein's existence is understood by way of thrownness-projection-fallenness.